\documentclass[12pt]{article}
\usepackage{lingmacros}
\usepackage[english]{babel}
\usepackage{graphicx}

\large \baselineskip = 20 true pt

\begin{document}
\begin{center}
{\Large \bf A Precedent Approach to Assigning Access Rights } \vspace{0.5cm}
\end{center}

\begin{center}
S.V. Belim, N.F. Bogachenko, A.N. Kabanov \\
Dostoevsky Omsk State University, Omsk, Russia
 \vspace{0.5cm}
\end{center}

\begin{center}
{\bf Abstract}
\end{center}

{\small
To design a discretionary access control policy, a technique is proposed that uses the principle 
of analogies and is based on both the properties of objects and the properties of subjects. 
As attributes characterizing these properties, the values of the security attributes of subjects
and objects are chosen. The concept of precedent is defined as an access rule explicitly 
specified by the security administrator. The problem of interpolation of the access matrix 
is formulated: the security administrator defines a sequence of precedents, it is required 
to automate the process of filling the remaining cells of the access matrix. On the family 
of sets of security attributes, a linear order is introduced. The principles of filling 
the access matrix on the basis of analogy with the dominant precedent in accordance with 
a given order relation are developed. The analysis of the proposed methodology is performed 
and its main advantages are revealed.
}

{\bf Keywords:} security model, discretionary access control, case based access control,
discretionary security policy.

\section{Introduction}

Discretionary access control policy (DAC) implies arbitrary access control: the administrator 
must define access rights for each subject-object pair. Obviously, even for small local systems,
the number of rules that need to be considered is estimated at tens of thousands. Such an access
control policy can not be set completely by the administrator. If we consider that the number of
objects and subjects in the computer system tends to increase, then the complexity of the task of
complete administration of the system exceeds practical possibilities.

A common approach for modern operating systems was the approach based on assigning default access
rights at the stage of object creation. The algorithm is that all objects created by a specific
subject are assigned the same access rights. In the future, if necessary, these rights can be
changed by the administrator. For example, in the Windows family of operating systems, 
the default rights are extracted from the context of the process that creates the file. 
This approach is based on the assumption that the same process works with objects that have 
the same security requirements. Thus, an approach based on the analog (precedent) method is used:
if some rights are set for one object created by this subject, then all objects created by this
subject should have similar access rights. The main drawback of this approach is that the analogy
is built exactly on one feature - the owner of the object.

In this article, we propose an algorithm for constructing DAC using the same analog (precedent)
method, but based on the properties of objects and on the properties of subjects. As parameters
characterizing the properties of objects and subjects, the values of their attributes are chosen.

Building an access control policy based on the attributes of files and processes was investigated
in a number of works. So, in the article \cite{b1} the HRU discretionary model was expanded 
and a typed access matrix was constructed, in which, in addition to object identifiers, their 
type is also used. The expansion of this model to the case of a dynamically changing object type
was carried out in \cite{b2}. These results were generalized taking into account not the type, 
but the attributes of the object in the ABAM model \cite{b3}. Also, the influence of attributes 
on role-based access policy was investigated \cite{b4,b5,b6,b7}. The influence of attributes 
on the authorization model was investigated in \cite{b8}. These works were developed as an
attribute-based access control model (ABAC). Access control standards based on the attributes of
subjects and objects were documented in NIST \cite{b5}. Unlike previous studies, we consider an
algorithmic approach to the formation of an access matrix and use the attributes of subjects and
objects as input data. Note that the approach implemented in the ABAC model is closer 
to role-based access differentiation, whereas the proposed algorithm remains within the
discretionary model.

\section{Sequence of precedents}

{\it 2.1. Access matrix}

When building a DAC, it is necessary to determine the access permission for each subject to each
object, starting from some formal rules that have a formal form.

Let $S$ be the set of subjects, $O$ be the set of objects, $P$ be the set of access rights. 
To define a DAC policy, it is necessary to set for each pair $(S_i, O_j)\in S\times O$ a certain
set of allowed access rights $\alpha\subseteq P$, that is, to define an access rule 
$(S_i, O_j,\alpha)$. Such rules are conveniently organized in a two-dimensional table of dimension $|S|\times |O|$, in which each subject has its own row, each object has its own column, 
and a set of allowed access rights $\alpha$ is indicated at their intersection. The constructed
table is usually called an access matrix, in what follows we denote it by $M$. The access matrix
defines a discrete map
\[
M: S\times O\rightarrow 2^P.
\]
The mapping $M$ is the decision function to allow or deny access and is called the access
function.

It should be noted that when the DAC policy is implemented in practice, for each "subject-object"
pair $(S_i, O_j)$ not an allowed set of access rights, but prohibited one may be specified. 
To distinguish these cases, we introduce the following notation:

1. access rule $(S_i, O_j,\alpha)$ ($[M]_{ij} =\alpha$) means that access $\alpha$ of the subject
$S_i$ to the object $O_j$ is allowed;

2. access rule $(S_i, O_j, \overline{\alpha})$ ($[M]_{ij}=\overline{\alpha}$) means that access
of the subject $S_i$ to the object $O_j$ is prohibited.

{\it 2.2. Security attributes}

{\bf Assumption 1.} We will assume that in the computer system for each subject and each object 
a set of properties is defined that characterizes the given subject or object and determines 
its rights in the system. Such a set of properties will be called security attributes.

You should not confuse security attributes with DAC access rights or security labels for 
a mandatory access policy. Security attributes are an integral part of the object 
and characterize its contents, type or status in the system. While the access rights are set 
in the system by the administrator quite arbitrarily. By entering the security attributes, 
we assume that the access rights to the objects with the same properties will be the same. 
It should be noted that not all object properties automatically refer to security attributes. 
Which attributes affect security, and which do not, determines the administrator, based on the
properties of the system as a whole.

Let each subject of system $S_i$ from set of subjects $S$ is characterized by a set of security
attributes $(a_1^i, ..., a_n^i)$, $a_k^i\in A_k$, $k = 1, ..., n$; and each object of system 
$O_j$ from set of objects $O$ is characterized by a set of security attributes 
$(b_1^j, ..., b_m^j)$, $b_s^j\in B_s$, $s = 1, ..., m$.

{\bf Assumption 2.} If two subjects have the same set of security attribute values, then they 
have the same access rights in the system.

{\bf Assumption 3.} If two objects have the same set of security attribute values, then 
the access rights to them of subjects with the same set of security attribute values are the same.

The last two assumptions introduce an equivalence relation on sets of subjects and objects. 
By equivalence, we mean sameness of two objects with the same set of values of security 
attributes of the security subsystem. In the future, all the subjects or objects, which are
identical in terms of security, we will refer to the same equivalence class, and a subject 
or object shall mean the corresponding equivalence classes. Working with equivalence classes
significantly reduces the complexity of checking the security of the system state. As is well
known, checking the security of an arbitrary system with DAC is not algorithmically solvable
\cite{b9}. The number of equivalence classes is always finite due to the limited number 
of security attributes and their values. Thus, each subject is a vector in $n$-dimensional
feature space: $S_i\in S\subseteq A_1\times ...\times A_n$, and the object is a vector 
in the m-dimensional feature space: $O_j\in O \subseteq B_1\times ...\times  B_m$. 	

Examples of security attributes of a subject include the following:

1. the user on whose behalf the subject is initialized;

2. the process level (kernel level or application level);

3. the location of the executable object, etc.

Examples of object security attributes:

1. system or not system object;

2. the type of the object;

3. the owner of the object;

4. the location of the object in the file system, etc.

{\it 2.3. Access rules}

{\bf Definition 1.} Let the security administrator explicitly fill in a certain cell 
of the access matrix, that is, the access rule $(S_i, O_j,\alpha)$ or 
$(S_i, O_j,\overline{\alpha})$ is defined, where $S_i\in S$, $O_j\in O$, 
$\alpha\subseteq P$ ($\alpha\neq \emptyset$). Such a triplet will be called a {\it precedent}
(or an {\it explicit access rule}).

The task is to set access rights, which are not explicitly defined by the system administrator, 
for "subject-object" pairs based on the analysis of the existing precedents. That is, it is
necessary to determine the values of the unknown access function M by some known set of its 
values. In this formulation, the problem reduces to the interpolation of a discrete function.

The problem of interpolation of the access matrix. Let subjects and objects be vectors in the
spaces of their attributes. Let the security administrator defines a sequence of precedents 
$Q=\{(S_{i1}, O_{j1},\alpha_1), ..., (S_{it}, O_{jt}, \alpha_t)\}$. It is required to fill 
in the remaining cells of the access matrix, that is, based on the set of given precedents,
interpolate the discrete access function.

{\bf Definition 2.} The access rule obtained in the automatic mode will be called an implicit
access rule.

In order to avoid contradictions in specifying the values of the cells of the access matrix, 
there should not be two precedents $q_a=(S_a, O_a,\alpha_a)$ and $q_b=(S_b,O_b,\alpha_b)$ 
in the sequence $Q$ such that 
$(S_a=S_b)\wedge (O_a=O_b)\wedge (\alpha_a\cap \alpha_b\neq \emptyset)$. Given that the
administration process is time-dispersed and the functioning of the computer system can lead 
to new requirements for access control, the condition on the sequence $Q$ may be violated. 
In this case, one of three approaches is possible.

1.	A new precedent is adopted and the old one is discarded. That is, it is considered that 
the administrator by default accepts only correct decisions and the new precedent corresponds 
to the changed requirements for the access control policy.

2.	A new precedent is not adopted. In this case, the access control policy does not change
significantly with each succeeding precedent. This approach is necessary in the administration 
of critical information processing systems.

3.	Interactive approach. In the event of a collision, the system in interactive mode asks 
the administrator which of the precedents, old or new, is considered correct.

In turn, the emergence of a new precedent leads to a redefinition of some of the implicit access
rules. In all cases, in addition to the full definition of the access matrix, the matrix $M$ is
determined ambiguously since it is possible to find various implicit access rules that satisfy 
the conditions imposed by a sequence of precedents.

\section{Interpolation of the access matrix}

{\it 3.1. Partial interpolation}

To determine implicit access rules, it is necessary to formulate some principles for their
building. First of all, we will be guided by the principle of issuing minimum rights, which 
is that when an uncertainty situation arises, the minimum of permissible sets of access rights 
is selected. However, if we confine ourselves to only one principle of minimal rights, we get 
a primitive solution, in which all accesses, except explicitly specified, are prohibited. 
Consider one of the possible examples of determining implicit access rules.

We assume that at the initial instant of time the access matrix is either filled, based on some 
a priori information, or, using the thesis "everything that is not allowed is forbidden," all 
the cells contain only access bans. We will fill the cells of the access matrix that correspond 
to implicit access rules in accordance with the following reasoning.

{\bf Definition 3.} A precedent that defines an implicit access rule is called dominant. 
If a dominant precedent $(S, O,\alpha)$ is found for the access of the subject $S_i$ 
to the object $O_j$, then the implicit access rule is determined by analogy with 
the dominating precedent as $(S_i, O_j,\alpha)$.

{\bf Assumption 4.} We assume that the subject's security attributes dominate the object's 
security attributes. Moreover, on the family of attribute sets we introduce a linear order: 
$A_1 > ... > A_n > B_1 > ... > B_m$.

A dominant precedent is chosen among the precedents influencing on this implicit access rule.
Selection of influencing and dominating precedents is carried out in accordance with the rules 
of partial interpolation of the access matrix:

1.	A precedent can influence on the access rights of subjects to objects only in its own row 
and in its own column of the access matrix.

2. The precedent $(S_{ip}, O_{jp},\alpha)$ influence on the access of the subject $S_i$ to the
object $O_j$, if:

$S_i = S_{ip}$ and for objects $O_j$ and $O_{jp}$ the values of at least one attribute 
are the same;

$O_j=O_{jp}$, and for subjects $S_i$ and $S_{ip}$ the values of at least one attribute 
are the same.

3.	If the access of the subject $S_i$ to the object $O_j$ is influenced by the precedents 
that specify both access of the subject $S_i$ (the precedents are located in the same 
row of the access matrix) and access to the object $O_j$ (the precedents are located in the 
same column of the access matrix), then, since the security attributes of the subject dominate 
the security attributes of the object, the precedents from the same row of the access matrix 
are more significant.

4.	If the access of the subject $S_i$ to the object $O_j$ is influenced by several precedents 
that determine the access of the subject $S_i$ (the precedents are located in the same row of 
the access matrix), then the linear order introduced on the set of security attributes of the
object is used to identify the dominant precedent: the precedent dominates with the more
significant coinciding attribute of objects.

5.	If the access of the subject $S_i$ to the object $O_j$ is influenced by several precedents 
that determine access to the object $O_j$ (the precedents are located in the same column of the
access matrix), then the linear order introduced on the set of security attributes of the subject
is used to identify the dominant precedent: the precedent dominates with more significant
coinciding attribute of subjects.

The specified rules for filling the access matrix allow us to draw the following conclusion. 
With partial interpolation, the order of precedents does not affect the resulting access matrix.
Indeed, the access of the subject $S_i$ to the object $O_j$ depends only on the precedents that
influence on it and does not depend on the previous state of the access matrix cells. 
The resulting algorithm for partial interpolation of the access matrix is formulated as follows:

1.	If the cell of the access matrix is defined by a precedent (an explicit access rule), its
contents remain unchanged.

2.	If the cell of the access matrix is not defined by the precedent, then an implicit access 
rule is formed for it based on the analogy (coincidence) with the dominant precedent. Selection 
of the dominant precedent occurs by comparing attributes in accordance with Assumption 4 and the
rules for partial interpolation of the access matrix.

{\bf Example 1.} Consider a subsystem that includes three subjects $S_1$, $S_2$, $S_3$ and three
objects $O_1$, $O_2$, $O_3$. We confine ourselves to two security attributes for the subjects 
of the system: $S\subseteq   A_1\times  A_2$. Objects will be characterized by three attributes: 
$O\subseteq  B_1\times  B_2\times B_3$. The values of the security attributes of subjects 
$S_1$, $S_2$, $S_3$ and objects $O_1$, $O_2$, $O_3$ are given as follows: 
$S_1 = (a_1^x, a_2^x)$, $S_2 = (a_1^y, a_2^x)$, $S_3 = (a_1^x, a_2^y)$, 
$O_1 = (b_1^x, b_2^x, b_3^x)$, $O_2 = (b_1^y, b_2^y, b_3^x)$, $O_3 = (b_1^x, b_2^y, b_3^z)$, 
with $a_1^x, a_1^y\in A_1$; $a_2^x, a_2^y\in A_2$; $b_1^x, b_1^y\in B_1$; $b_2^x, b_2^y\in B_2$;
$b_3^x, b_3^z\in B_3$. Suppose that $P = \{all\}$. Then $[M]_{ij} = 1$ if full access is allowed,
and $[M]_{ij} = 0$ if full access is denied. Access permissions or access denials defined by
precedents will be marked in square brackets in the access matrix. The default accesses will 
be denoted by the sign "?".

Suppose that only one precedent is created: $q_1 = (S_1, O_1, 1)$. Consider the accesses of the
same subject to the two remaining objects. Both access $S_1$ to $O_2$ and $S_1$ to $O_3$ will 
be allowed since the object $O_2$ coincides with $O_1$ by the third attribute $b_3^x$, 
and $O_3$ coincides with $O_1$ by the first attribute b1x. The access rights of subjects $S_2$ 
and $S_3$ to the object $O_1$ will be the same as for the subject $S_1$ since $S_1$ and $S_2$ 
have the same second attribute $a_2^x$, and for $S_1$ and $S_3$ the first attribute a1x coincides
(see Table 1).

\begin{table}
\caption{Partial interpolation of the access matrix for $Q = \{q_1\}$.}
\begin{center}
\begin{tabular}{cccc}
  \hline
 &$O_1 = (b_1^x, b_2^x, b_3^x)$ & $O_2 = (b_1^y, b_2^y, b_3^x)$ & $O_3 = (b_1^x, b_2^y, b_3^z)$ \\
 \hline
  $S_1 = (a_1^x, a_2^x)$ & [1] & 1 &1\\
  $S_2 = (a_1^y, a_2^x)$ & 1 & ? & ? \\
  $S_3 = (a_1^x, a_2^y)$ & 1 & ? & ?\\
  \hline
\end{tabular}
\end{center}
\end{table}

Now suppose that the precedent $q_2 = (S_1, O_3, 0)$ has occurred in the system. The permission
to access $S_1$ to $O_2$ is not explicitly specified. The first precedent allows this access,
because the third attribute $b_3^x$ is the same, and the second precedent prohibits it, because 
the second attribute $b_2^y$ is the same. But $B_2 > B_3$, that is, the second attribute is more
significant than the third one, so access of $S_1$ to $O_2$ will be prohibited. The access rights
of subjects $S_2$ and $S_3$ to the object $O_3$ will be the same as for the subject $S_1$ since 
$S_1$ and $S_2$ have the same second attribute $a_2^x$, and for $S_1$ and $S_3$ the first 
attribute $a_1^x$ coincides (see Table 2).

\begin{table}
\caption{Partial interpolation of the access matrix for $Q = \{q_1, q_2\}$.}
\begin{center}
\begin{tabular}{cccc}
  \hline
 &$O_1 = (b_1^x, b_2^x, b_3^x)$ & $O_2 = (b_1^y, b_2^y, b_3^x)$ & $O_3 = (b_1^x, b_2^y, b_3^z)$ \\
 \hline
  $S_1 = (a_1^x, a_2^x)$ & [1] & 0 &[0]\\
  $S_2 = (a_1^y, a_2^x)$ & 1 & ? & 0 \\
  $S_3 = (a_1^x, a_2^y)$ & 1 & ? & 0\\
  \hline
\end{tabular}
\end{center}
\end{table}

Let an additional precedent $q_3 = (S_2, O_2, 1)$ be created. Consider access $S_2$ to $O_3$. 
This access will be allowed since objects have the same attribute $b_2^y$, and the precedent 
$q_3$ in this case dominates the precedent $q_2$ since the attributes of the subject dominate 
the attributes of the object. Similarly, the precedent $q_3$ defines the access of the subject 
$S_2$ to the object $O_1$ and does not influence on the access of the subject $S_1$ to the object
$O_2$. The precedent $q_3$ does not influence on the access of $S_3$ to $O_2$ since subjects 
$S_2$ and $S_3$ have no coinciding attributes (see Table 3).

\begin{table}
\caption{Partial interpolation of the access matrix for $Q = \{q_1, q_2, q_3\}$.}
\begin{center}
\begin{tabular}{cccc}
  \hline
 &$O_1 = (b_1^x, b_2^x, b_3^x)$ & $O_2 = (b_1^y, b_2^y, b_3^x)$ & $O_3 = (b_1^x, b_2^y, b_3^z)$ \\
 \hline
  $S_1 = (a_1^x, a_2^x)$ & [1] & 1 &1\\
  $S_2 = (a_1^y, a_2^x)$ & 1 & [1] & 1 \\
  $S_3 = (a_1^x, a_2^y)$ & 1 & ? & 0\\
  \hline
\end{tabular}
\end{center}
\end{table}

Once again, we note that implicit access rules do not depend on the order of precedents. 
So, in the example considered, to obtain the resultant access matrix (Table 3), it is sufficient 
to know that there were precedents $q_1$, $q_2$, $q_3$.

{\it 3.2. Sequential interpolation}

Are there other ways of interpolating the access matrix? We require that the precedent can
influence the access rights of subjects to objects not only in its own row and in its own column 
of the matrix. That is, the interpolation algorithm of the access matrix must be a "chain
reaction". Assuming that each precedent potentially influences on the entire access matrix, 
the following rules for sequential interpolation of the access matrix can be proposed:

1.	Each new precedent can change the accesses in the "own" row. Then accesses in the "own" 
column can change. That is, the rules of partial interpolation described in the previous 
section apply to the first stage of processing the precedent.

2.	At the second stage, each cell from the precedent's row that changed its state is considered 
a precedent for the cells of its column. Here the rules of partial interpolation in the column
again apply.

It should be noted that, as in the case of partial interpolation, the resulting access matrix 
does not depend on the order of precedents.

{\bf Example 2.} Tables 4, 5, and 6 show the result of the rules of sequential interpolation 
of the access matrix. Angular brackets denote accesses in the cells of the precedent's row that
have changed their state. For example, in Table 5, access of $S_1$ to $O_2$ is determined by the
use of precedent $q_2$, and accesses of $S_2$ to $O_2$ and $S_3$ to $O_2$ are determined under 
the influence of access $S_1$ to $O_2$.

\begin{table}
\caption{Sequential interpolation of the access matrix for $Q = \{q_1\}$.}
\begin{center}
\begin{tabular}{cccc}
  \hline
 &$O_1 = (b_1^x, b_2^x, b_3^x)$ & $O_2 = (b_1^y, b_2^y, b_3^x)$ & $O_3 = (b_1^x, b_2^y, b_3^z)$ \\
 \hline
  $S_1 = (a_1^x, a_2^x)$ & [1] & $>1<$ &$>1<$\\
  $S_2 = (a_1^y, a_2^x)$ & 1 & 1 & 1 \\
  $S_3 = (a_1^x, a_2^y)$ & 1 & 1 & 1\\
  \hline
\end{tabular}
\end{center}
\end{table}

\begin{table}
\caption{Sequential interpolation of the access matrix for $Q = \{q_1, q_2\}$.}
\begin{center}
\begin{tabular}{cccc}
  \hline
 &$O_1 = (b_1^x, b_2^x, b_3^x)$ & $O_2 = (b_1^y, b_2^y, b_3^x)$ & $O_3 = (b_1^x, b_2^y, b_3^z)$ \\
 \hline
  $S_1 = (a_1^x, a_2^x)$ & [1] & $>0<$ &[0]\\
  $S_2 = (a_1^y, a_2^x)$ & 1 & 0 & 0 \\
  $S_3 = (a_1^x, a_2^y)$ & 1 & 0 & 0\\
  \hline
\end{tabular}
\end{center}
\end{table}

\begin{table}
\caption{Sequential interpolation of the access matrix for $Q = \{q_1,q_2, q_3\}$.}
\begin{center}
\begin{tabular}{cccc}
  \hline
 &$O_1 = (b_1^x, b_2^x, b_3^x)$ & $O_2 = (b_1^y, b_2^y, b_3^x)$ & $O_3 = (b_1^x, b_2^y, b_3^z)$ \\
 \hline
  $S_1 = (a_1^x, a_2^x)$ & [1] & 0 &[0]\\
  $S_2 = (a_1^y, a_2^x)$ & 1 & [1] & $>1<$ \\
  $S_3 = (a_1^x, a_2^y)$ & 1 & 0 & 0\\
  \hline
\end{tabular}
\end{center}
\end{table}

{\it 3.3. Uncertainty situations}

For the presented algorithms of interpolation of the access matrix, there can be situations 
when existing precedents do not allow to determine the type of access for a certain
"subject-object" pair. This happens in the following situations:

there are no precedents influencing on access (see access of subject $S_1$ to object $O_5$ 
in Table 7);

from precedents that influence on access you cannot select the dominant one (see access of 
subject $S_1$ to object $O_7$ in Table 7).

\begin{table}
\caption{Interpolation of the access matrix for $Q = \{(S_1, O_4, 1), (S_1, O_6, 0), (S_1, O_8, 0)\}$.}
\begin{center}
\begin{tabular}{cccccc}
  \hline
  &$O_4 = (b_1^x, b_2^x)$	&$O_5 = (b_1^y, b_2^y)$	&$O_6 = (b_1^z, b_2^z)$	&$O_7 = (b_1^x, b_2^s)$	
  &$O8 = (b_1^x, b_2^t)$\\
 \hline
  $S_1 = (a_1^x, a_2^x)$	&[1]	&?	&[0]	&?	&[0]\\
  \hline
\end{tabular}
\end{center}
\end{table}

\section*{Conclusion}

The use of additional security attributes of subjects and objects, determined by the system
administrator, significantly expands the options for configuring the DAC policy. The
precedent-based approach can be seen as the development and improvement of the default access
control system. At the same time, the proposed method of administration the DAC policy refers to
decision support methods - algorithms allow to partially automate the process of assigning access
rights and provide the administrator with information about access that cannot be determined
automatically and require an explicit task.

The developed algorithms do not depend on the order of precedents, which allows you to
significantly reduce the amount of memory to store data on the behavior of the system.

It is easy to show that the algorithms for partial and sequential interpolation of the access
matrix are polynomial with respect to the quantitative characteristics of the system. Indeed, the
implementation of the rules of partial interpolation (rules of the precedent's influence in the 
row and in the column) will require no more than $O(m|O|^2 + n|S|^2)$ operations. The complexity 
of the second stage of the sequential interpolation algorithm does not exceed 
$O(n\cdot|O|\cdot |S|^2)$. Since the entities of the system are represented by equivalence 
classes defined by finite sets of values of security attributes, the software implementation 
of the algorithms for interpolating the access matrix is not difficult.

In the examples given, we limited ourselves to considering only full access. The proposed
algorithms can easily be extended to any number of possible accesses. In this case, 
each precedent specifies its own set of accesses, which either replaces, or does not, 
the type of access in the cells of the matrix influenced by the precedent.

The use of this approach to the formation of an access matrix may prove to be productive 
in distributed systems for which there is a problem of reconciling data at different nodes. 
When changing the access rules, the administrator does not need to forward the entire 
new access matrix, it is enough to send out precedents, on the basis of which all nodes will 
form access permissions using the same algorithm. Forwarding individual precedents reduces 
system performance requirements.

Productivity can also be enhanced by more "soft" formation of access permission when creating new
objects and subjects. It is enough for the administrator to set the correct values 
for the security attributes of the object and the system will independently create access control.

\end{document}